\documentstyle[epsfig,12pt]{article}
\begin{document}

\begin{center}

{\bf
\title "The equilibrium of the dense electron-nuclear plasma in the gravitational field.
The magnetic fields and masses of stars.}

\bigskip

\bigskip
\author "B.V.Vasiliev
\bigskip

Institute in Physical-Technical Problems, 141980, Dubna, Russia
\bigskip

{vasiliev@dubna.ru}
\end{center}

\bigskip
%\maketitle

\begin{abstract}
The equilibrium of a hot dense plasma in a gravitational field is
considered. From the standard equilibrium equations, the energy
minimum at density about $10^{25}$ particles per $cm^3$ and
temperature about $10^7 K$ was found. This effect plays an
important role for astrophysics. It enables to explain
the mechanism of the star magnetic field generation and to make a
prediction for the spectrum of a star mass with a wholly
satisfactory agreement for the observation data.
\end{abstract}
\bigskip

\bigskip

\bigskip
\bigskip

\section{The current hypothesis}

    Now it is conventionally accepted to think that a density and
temperature of a star interior substance are growing depthward of
stars and can amount to  tremendous values at their central cores.
It seems that this growing is a necessary condition of an
equilibrium of a star substance in the self gravity field.

    The substance exists as a hot dense electron-nuclear plasma at
high pressure and temperature of a star interior. At this
condition in zero approximation, plasma can be considered as a
Boltzmann ideal gas with energy
\begin{equation}
E=\frac{3}{2}kT N \label{eB}
\end{equation}
where $N$ is the particle number.

As the a direct inter-nuclear interaction is small in plasma, it
can be neglected and one can write the equilibrium equation in the
form \cite{1}:

\begin{equation}
\mu_e+m'\psi=const\label{equ1}
\end{equation}
where $\mu_e$ is the electron chemical potential,
$m'={\gamma}/{n_e}$ is the mass of substance related to one
electron, $n_e$ is the electron gas density, $\gamma$ is the mass
density of plasma, $\psi$ is the Newton gravitational potential.
Because $\Delta \psi=-4\pi G \gamma$, in a spherically symmetric
case, it is reduced to

\begin{equation}
\frac{1}{r^2} \frac{d}{dr}\biggl(r^2 \frac{d\mu_e}{dr}\biggr)
=-4G\pi\gamma m'.\label{equ2}
\end{equation}
The chemical potential of electron gas in the Boltzmann case
\cite{1}
\begin{equation}
\mu_e= kT ln \biggl[\frac{n_e}{2}\biggl(\frac{2\pi \hbar^2}{m
kT}\biggr)^{3/2}\biggr]\label{muB}
\end{equation}
is the function of particle density and temperature only. A
single-meaning conclusion can be deduced from the equilibrium
equation (Eq.({\ref{equ2}})) if to consider hot non-degenerate
plasma as an ideal gas: the balance of plasma particles in the
self gravity field demands a temperature and density increasing
depthward of a star.

The consideration of a hot dense electron-nuclear plasma as  ideal
gas is possible in zero approximation only because it means that
the inter-particle interaction is completely neglected by
definition. At first sight it is acceptable as the inter-particle
interaction are small in comparison with the ideal gas energy
(Eq.({\ref{eB}})). But the allowance of this interaction has a
principal importance because it forms a stable equilibrium state
of a hot plasma. It is lost at the consideration of plasma in the
ideal gas approximation.

\section{The density and temperature of a hot dense plasmas in the equilibrium state}

\subsection{The steady-state density of a hot dense plasma}

There are two main characteristic features which we must take to
account at hot dense plasma consideration. The first of them is
related to the quantum  properties of  electron gas. The
second feature is concerned with the presence of positively
charged nuclei inside an electron gas of plasma.

\subsubsection{The correction for Fermi-statistics}

The estimation of  electron gas energy in the Boltzmann case
$(kT\gg E_F)$ can be obtained by an expansion in series of the
non-relativistic Fermi-particle system full energy \cite{1}:

\begin{equation}
E=\frac{2^{1/2}V m_e^{3/2}}{\pi^2 \hbar^3} \int_0^\infty
\frac{\varepsilon^{3/2}d\varepsilon}
{e^{(\varepsilon-\mu_e)/kT}+1}. \label{eg}
\end{equation}

Where $\varepsilon$ is energy of particle. Because, in the
Boltzmann case, $\mu_e<0$ and $|\mu_e/kT|\gg 1$, the integrand at
$e^{\mu_e/kT}\ll 1$ can be expanded into a series according to
their powers $e^{\mu_e/kT-\varepsilon/kT}$.

As a result, the full energy of a hot electron gas \cite{1} with
account of its quantum properties is
\begin{equation}
E=\frac{3}{2} kT
\biggl[1+\frac{\pi^{3/2}}{4}\biggl(\frac{a_0e^2}{kT}\biggl)^{3/2}n_e\biggr],
\label{cef}
\end{equation}
where $a_0=\frac{\hbar^2}{me^2}$ is Bohr radius.

It is important to underline that the correction for
Fermi-statistic is positive, because it takes into account that
electron can not take places which are occupied by other electrons
and a resulting pressure is more than pressure of an ideal gas at
just the same density and temperature.

\subsubsection {The correction for a correlation of charged
particles in plasma}

At high temperature, the plasma particles approach to an uniform
space distribution. At this limit, the energy of ion and electron
interaction is equal to zero. Some correlation in space
distribution of particles arises as much as the positively charged
particle groups  itself preferably around particles with negative
charges and vice versa. It is accepted to estimate the energy of
this correlation by the method developed by Debye-H$\ddot{u}$kkel
for strong electrolytes \cite{1}. The energy of a charged particle
inside plasma is equal to $e\varphi$, where $e$ is a charge of a
particle, and $\varphi$ is the electric potential induced by other
particles  on the particle under consideration.

This potential inside plasma is determined by the Debye law
\cite{1}:

\begin{equation}
\varphi(r)=\frac{e}{r} e^{-\frac{r}{r_D}}\label{vr}
\end{equation}
where the Debye radius is

\begin{equation}
r_D=\sqrt{\frac{kT}{4\pi e^2 n_e}}\label{rD}
\end{equation}
For small values of ratio $\frac{r}{r_D}$, the potential can be
expanded into a series

\begin{equation}
\varphi(r)=\frac{e}{r}-\frac{e}{r_D}+...\label{rr}
\end{equation}
The following terms are converted into zero at $r\rightarrow 0$.
The first term of this series is the potential of the considered
particle. The second term is a potential induced by other
particles of plasma on the charge under consideration. Therefore,
the correlation energy of plasma is \cite{1}

\begin{eqnarray}
\delta E_{corr}=-N_e\frac{\pi^{1/2}e^3
kT}{n_e}\biggl[Z^2\frac{n_n}{kT} +\biggl(\frac{\partial
n_e}{\partial \mu_e} \biggr)_{N,T}\biggr]^{3/2},\label{Fcor}
\end{eqnarray}
where $Z$ is the nuclear charge and the density of nuclei is
$n_n=\frac{n_e}{Z}$.

Because the chemical potential of the Boltzmann ideal gas is
determined by (Eq.{\ref{muB}}), at high temperature

\begin{equation}
\frac{d\mu_e}{dn_e}=\frac{kT}{n_e}
\end{equation}
and

\begin{equation}
\delta E_{corr}=-N_e \biggl(\frac{\pi
n_e}{kT}\biggr)^{1/2}(Z+1)^{3/2}e^3
\end{equation}
As an attraction between unlike charges which  are placed nearer
one to another is prevalent over a repulsion of like charges, the
plasma pressure is below  the pressure of the ideal gas at the same parameters.
 For this reason this correction is a
negative one.

\subsubsection{The density of a hot dense plasma at the equilibrium state}

Thereby the full energy of plasma with account of the main
corrections on inter-particle interaction is

\begin{equation}
E=\frac{3}{2}kTN_e\biggl[1+\frac{\pi^{3/2}}{4}\biggl(\frac{a_0
e^2}{kT}\biggr)^{3/2}n_e - \frac{2\pi
^{1/2}}{3}\biggl(\frac{Z+1}{kT}\biggr)^{3/2}e^3 n_e^{1/2}\biggr]
\end{equation}

At a constant full number of particles in the system and at
constant temperature, the condition of the equilibrium state
existence at the energy minimum is

\begin{equation}
\biggl(\frac{\partial E}{\partial n_e}\biggr)_{N,T}=0,\label{dedn}
\end{equation}
what allows one to obtain the steady-state value of density of hot
non-relativistic plasma

\begin{equation}
n_{\star}=\frac{16(Z+1)^{3}}{9 \pi^2 a_0^3}\simeq 2\cdot
10^{24}(Z+1)^3 ~ cm^{-3}, \label{neq}
\end{equation}
Here the Fermi-energy of electron gas of equilibrium plasma at
this density is

\begin{equation}
\varepsilon_F(n_{\star})= \biggl(\frac{16}{3}\biggr)^{2/3}\frac{m
e^4}{2\hbar^2}(Z+1)^2\approx 1.5~\frac{e^2}{a_0}(Z+1)^2\label{eF}
\end{equation}

\subsection{Equilibrium temperature of a hot non-relativistic
star} As the steady-state value of density of hot non-relativistic
plasma is known, we can obtain a steady-state value of temperature
of a hot non-relativistic plasma.

According to the virial theorem \cite{1,BV+Lub}, the potential
energy $U$ of particles with Coulomb interaction is equal to
 their double kinetic energy $T$ with an opposite sign

\begin{equation}
U =-2T
\end{equation}
and their full energy is equal to  kinetic energy with an opposite
sign. At neglecting of small corrections at high temperature, one
can write the full energy of hot dense plasma as

\begin{equation}
E_{plasma}= U + \frac{3}{2}kTN = - \frac{3}{2}kTN.
\end{equation}
As the plasma temperature is high enough, pressure of black
radiation cannot be neglected. The full energy of a star depending
on the plasma energy and the black radiation energy is

\begin{equation}
E_{total}=-\frac{3}{2}kTN + \frac{\pi^2}{15}
\biggl(\frac{kT}{\hbar c}\biggr)^3 V kT
\end{equation}
The equilibrium temperature of a body consisting of hot
non-relativistic plasma is determined by the energy minimum
condition

\begin{equation}
\biggl(\frac{\partial E_{total}}{\partial T}\biggr)_{N,V} =0.
\end{equation}
It gives

\begin{equation}
T_{\star}=\biggl(\frac{10}{\pi^4}\biggr)^{1/3}(Z+1)\frac{\hbar
c}{ka_0}\approx 2\cdot (Z+1)\cdot 10^7~K.\label{4a}
\end{equation}

Usually all substances have a positive thermal capacity. Therefore
the minimal energy for such substances exists at $T=0$. The
existence, in our case, of the energy minimum at finite
temperature $T_{\star}\neq 0$ is not confusing. Each small part of
a star has a positive thermal capacity, but a gravitational
interaction of these parts between themselves results into that
the thermal capacity of a star as the whole  becomes negative at
some temperature and a star energy decreases with an increased
temperature. As a result there are two branches of the temperature
dependence of a star energy - with the negative capacity at low
temperature and with a positive capacity at high temperature.
Between them, at some finite temperature $T_{\star}$ there is a
minimum of energy.

Steady-state values of density and temperature for hot
non-relativistic plasma have been considered above. One can see, that the
criterion according to which plasma may be considered as a hot one

\begin{equation}
kT \gg \varepsilon_F
\end{equation}
with account of Eq.{\ref{eF}} is satisfied if the nuclear charge
is not too big

\begin{equation}
\frac{kT_{\star}}{\varepsilon_F(n_{\star})}=\biggl(\frac{45}{16\pi^4}\biggr)^{1/3}~\frac{\hbar
c}{(Z+1)e^2} \approx \frac{0.3}{Z+1}~\alpha^{-1}\approx
\frac{40}{Z+1},
\end{equation}
where $\alpha=e^2/\hbar c=1/137$ is the fine structure constant.

\section{The equilibrium of a dense electron-nuclear
plasma}

According to definition (Eq.({\ref{muB}})) in plasma equilibrium
state at a constant temperature and density, its chemical
potential must be constant too:

\begin{equation}
\mu(n_{\star},T_{\star})=const.
\end{equation}

(One can note, that for a cold degenerate plasma the chemical
potential does not depend on temperature and it is constant at
$n=const$.)

Thus the equilibrium of plasma in gravity field can be obtained
if  the role of the field of another nature than
gravitational one,  for example, the electric field, is taken into account. Speaking more precisely
 the equilibrium equation (Eq.({\ref{equ1}})) must imply
all fields, which can have an impact on particles, for example,
electric field for system of charged particles:
\begin{equation}
\mu+\sum_i q_i\phi_i=const
\end{equation}
(where $q_i$ and $\phi_i$ are charge and potential of any nature
(gravitational,electric)).

In a spherical symmetric case at $\mu=const$ it reduces to

\begin{equation}
G\gamma m'=\rho q.
\end{equation}

where $\rho=q\cdot n_e$ is the electric charge density,
\begin{equation}
q=G^{1/2}m'\label{q}
\end{equation}
is the charge induced in plasma cell by the gravity (related to one
electron). One should not think that the  gravity field really
induces in plasma some additional charge. We can rather  speak about
electric polarization of plasma which can be described as some
redistribution of internal charges in the plasma body. Essentially
it stays electrically neutral as a whole, because the positive
charge with volume density

\begin{equation}
\rho=G^{1/2}\frac{\gamma}{Z}
\end{equation}
is concentrated inside the charged plasma core and the
corresponding negative electric charge exists on its surface.

Thus, in a spherical symmetric case the equilibrium equation can
be rewritten as

\begin{equation}
\gamma \mathbf{g}+\rho \mathbf{E}= \mathrm{0}\label{f1}
\end{equation}
where

\begin{equation}
{\mathbf{E}}=\frac{{\mathbf{g}}}{(GZ)^{1/2}}.
\end{equation}

The Thomas-Fermi approximation permits to consider the balance in
plasma cells in more detail \cite{BV}.

\subsubsection{The equilibrium density of another kind of dense plasmas}

The above consideration of equilibrium of hot non-relativistic
plasma is obvious, but  it is a characteristic property not of  this
plasma only.

The direct consideration of a plasma equilibrium in Fermi-Thomas
approximation shows that the application of a gravity field to
plasma induces its electric polarization
(\cite{BV},\cite{BV-book}) and so the equilibrium equation in the
form Eq.({\ref{f1}}) is applicable to all kind of dense plasmas -
relativistic or non-relativistic and simultaneously degenerate or
non-degenerate ones.

Essentially at that one can find  the constancy of density and
chemical potential of another kind of plasmas in an equilibrium
condition.

So for a cold non-relativistic plasma the kinetic energy of electron
is

\begin{equation}
E_k=\frac{3}{5}E_F=\frac{3}{10} (3\pi^2)^{2/3}a_0 e^2 n_e^{2/3}.
\end{equation}

Its potential energy is

\begin{equation}
E_p\approx - e^2 n_e^{1/3}.
\end{equation}
As according to the virial theorem $E_k\approx -E_p$ the
equilibrium electron density is

\begin{equation}
n_e \approx a_0^{-3}.
\end{equation}
and it does not depend on temperature.

The relativistic plasma exists at huge pressure which induces the
neutronization of substance. For this process density

\begin{equation}
n_e=\frac{\triangle^{3}}{3\pi^2 (c\hbar)^3} \label{neu}
\end{equation}
is characteristic \cite{1}. Where $\triangle$ is the difference of
nuclear bonding energy of neighbouring interacting nuclei. Because of
this difference

\begin{equation}
\triangle \approx m_e c^2\label{ne1}
\end{equation}
the equilibrium of relativistic plasma density (at condition of a
homogeneous mixing of reacting substance)

\begin{equation}
n_e\approx \frac{1}{3\pi^2}(\alpha a_0)^{-3} \approx 10^{30}
cm^{-3}\label{ne2}
\end{equation}

Where $\alpha=e^2/\hbar c$ is the fine structure constant,
$a_0=\hbar^2/m_ee^2$ is the Bohr radius.

A similar consideration can be extended on the neutron matter if
 it is considered as electron-proton plasma in the neutron
environment.

\section{The giro-magnetic ratio of stars}

Gravitation produces a redistribution of free charges in
plasma inside a star. Essentially, a star as a whole conserves its
electric neutrality. However, as the star rotates about its axis,
positive volume charges are moving on smaller radial distances
than the surface negative charge. It induces a magnetic field
which can be measured.

The magnetic moment of surface spherical layer, which carries the
charge $Q$, is

\begin{equation}
\mu_-=-\frac{1}{3}Q\Omega R^2
\end{equation}

Where $\Omega$ is rotational velocity.

The magnetic moment induced by a volume charge is

\begin{equation}
\mu_+=\frac{4\pi\Omega}{3} \int_0^R \rho r^4dr =\frac{1}{5}Q\Omega
R^2
\end{equation}

Where  $Q=\frac{4\pi}{3}\rho R^3$.

Thus, the summary magnetic moment of a star is

\begin{equation}
\mu_{\Sigma}=-\frac{2}{15}\biggl(\frac{4\pi}{3}\rho R^3\biggr)
\Omega R^3
\end{equation}

As the mass density inside a star is constant, the angular
momentum of a star is

\begin{equation}
L=\frac{2}{5}M\Omega R^2
\end{equation}

According to Eq.({\ref{q}}), we obtain that the giro-magnetic
ratio is expressed through world constants only:

\begin{equation}
\vartheta=\frac{\mu_{\Sigma}}{L}=\frac{\sqrt{G}}{3c}\label{3a}
\end{equation}

It can be verified by measurement data.

\begin{figure}
\begin{center}
\includegraphics[8cm,9cm][14cm,15cm]{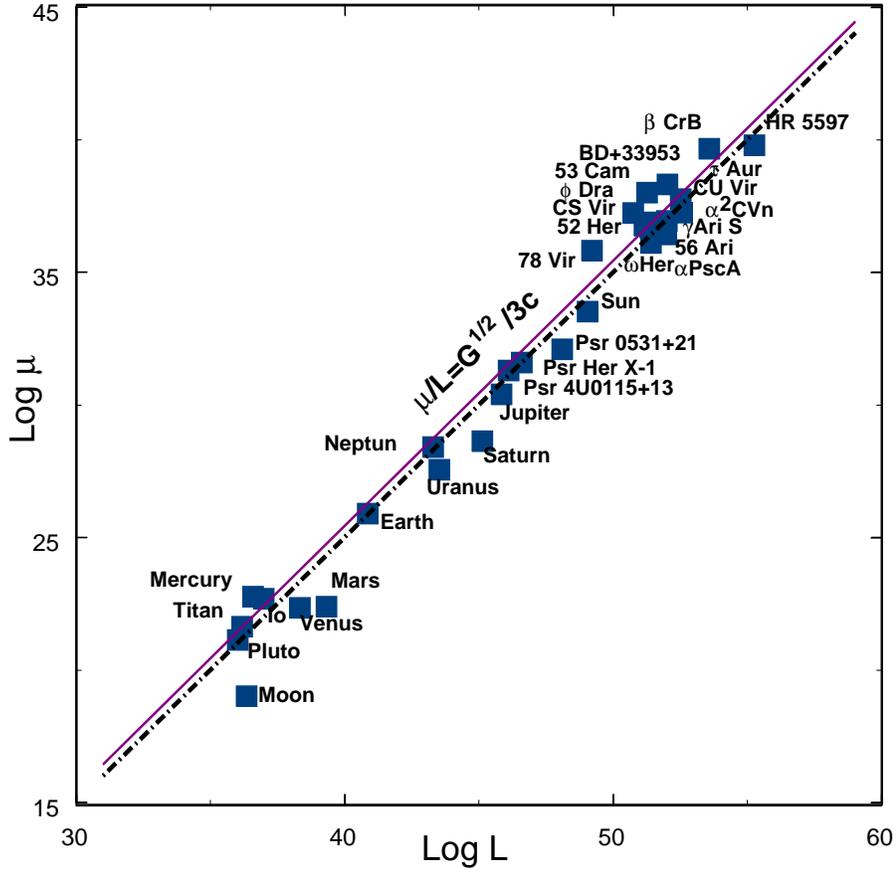}
\vspace{6cm} \caption { The observed values of the magnetic
moments of celestial bodies vs. their angular momenta. On the
ordinate, the logarithm of the magnetic moment over
$Gs\cdot{cm^3}$ is plotted; on the abscissa the logarithm of the
angular momentum over $erg\cdot{s}$ is shown. The solid line
illustrates  Eq.({\ref{3a}}). The dash-dotted line is the fitting
of the observed values.} \label{mL}
\end{center}
\end{figure}
%\clearpage

The values of giro-magnetic ratio for all celestial bodies (for
which they are known today) are shown in Fig.{\ref{mL}}.

The data for planets are taken from \cite{Sirag}, the data for
stars are taken from \cite{Borra}, and  those for pulsars - from
\cite{Beskin}. Therefore, for all celestial bodies - for planets
and their satellites, for $Ap$-stars and several pulsars - the
calculated value of gyromagnetic ratio Eq.({\ref{3a}}) with a
logarithmic accuracy quite satisfactorily agrees with
measurements, which angular momenta themselves change within the
limits of more than 20 orders.

\section{The stellar mass distribution}
\subsection{The mass of star consisting of hot non-relativistic dense plasma}

Inside the stellar core consisting of hot dense plasma the
gravitational force is counterbalanced by the electric force
(Eq.({\ref{f1}})) and gradient of pressure is absent. The absence
of a pressure gradient inside the hot star core does not mean
absence of pressure. It is not  difficult to make sure, if the
gravity force inside a star is compensated by electric force, the
negative energy of gravitational field is cancelled by the energy
of electric field. The non-compensated part of energy is the
energy of gravitational field outside a star. This field has an
energy

\begin{equation}
E_G=-\frac{GM_{\star}^2}{2R_{\star}}
\end{equation}
where $M_{\star}$ and $R_{\star}$ are the mass and radius of a hot
plasma core. This external gravitational field tends to compress a
star.

By definition, pressure of a gas (at entropy S=0) is

\begin{equation}
P=-\biggl(\frac{\partial E_G}{\partial V}\biggr)_{s=0}=
-\frac{{E_G}}{3V_{\star}}=\frac{GM_{\star}^2}{6R_{\star}V_{\star}},\label{3b}
\end{equation}
where $V_{\star}=\frac{4\pi}{3}R_{\star}^3$ is the volume of a
star.

 The equilibrium of a star exists when the gravitational
pressure, which tends to compress a star, is counterbalanced by
internal pressure of the hot electron gas and by pressure of the
radiation:

\begin{equation}
\frac{GM_{\star}^2}{6RV}=kT_{\star}n_{\star} + \frac{\pi^2}{45}
\frac{(kT_{\star})^4}{(\hbar c)^3}
\end{equation}

Finally we have the mass of a star core consisting from hot dense
plasma
\begin{equation}
M_{\star}=1.5^6 \sqrt{\frac{10}{\pi^3}} \biggl(\frac{\hbar c}
{Gm_p^2}\biggr)^{3/2} \biggl(\frac{Z}{A}\biggr)^2 m_p\nonumber\\
\approx  \frac{6.47M_{Ch}}{(A/Z)^2} \approx
\frac{12.0M_{\odot}}{(A/Z)^2}, \label{4b}
\end{equation}
where $M_{Ch}=\biggl(\frac{\hbar c}{G
m_p^2}\biggr)^{3/2}m_p=3.42\cdot 10^{33} g$ is the Chandrasechar
mass. It is important to underline that the obtained stellar mass
estimation Eq.({\ref{4b}}) is depending on only one parameter
$A/Z$.

One can note that the equilibrium radius of star core

\begin{eqnarray}
R_{\star}= \frac{(3/2)^3}{2}
\biggl(\frac{10}{\pi}\biggr)^{1/6}\biggl(\frac{\hbar c}{G
m_p^2}\biggr)^{1/2} \frac{a_0}{(Z+1){A/Z}}.\label{s10}
\end{eqnarray}

such as depends on $A$ and $Z$ only.

\subsection{The mass of star consisting of cold relativistic plasma}

With an increase of density, the plasma can turn into
relativistic state. It occurs when Fermi momentum of electrons

\begin{equation}
p_F=(3\pi^2)^{1/3}n_e^{1/3} \hbar>m_e c
\end{equation}

%$n\approx 10^{31} 1/cm^3$

This value of momentum corresponds to the steady-state density of
a substance under neutronization (Eq.({\ref{ne2}})) at
$n_{\ast}\approx 10^{30} ~cm^{-3}$. At  temperature

\begin{equation}
T\ll\frac {mc^2} {k} \approx 10^{10}~K.
\end{equation}
it can be considered as cold.

As the pressure of a relativistic electron gas is
\begin {equation}
P_R=\frac{(3\pi^2)^{1/3}}{4} n^{4/3}\hbar c,
\end {equation}
in accordance with Eq.({\ref{3b}}), the pressure balance obtains
the form:

\begin {equation}
\frac{GM_{\ast}^2}{6R_{\ast}V_{\ast}}=\frac{(3\pi^2)^{1/3}}{4}
n^{4/3}\hbar c,
\end {equation}

Therefore, the relativistic degenerate star in equilibrium state
must have a steady value of mass

\begin {equation}
M_{\ast}= {1.5}^{5/2}\pi^{1/2}\biggl(\frac{\hbar c}{Gm_p^2}
\biggr)^{3/2}\frac{m_p}{(A/Z)^2}\approx
\frac{4.9M_{Ch}}{(A/Z)^2}\approx
\frac{9.0M_{\odot}}{(A/Z)^2}\label{5b}
\end {equation}

at radius corresponding to Eq.({\ref{ne2}})

\begin {equation}
R_{\ast}\approx\biggl(\frac{\hbar c} {G m_p^2}\biggr)^{1/2}
\frac{\alpha a_0}{A/Z}\approx \frac{10^{-2}R_{\odot}}{(A/Z)}
\end {equation}

The objects which have such masses and density are best suited to
the dwarfs.

\subsection{The comparison of calculated star masses with observations}

\begin{figure}
\begin{center}
\includegraphics[10cm,3cm][9cm,12cm]{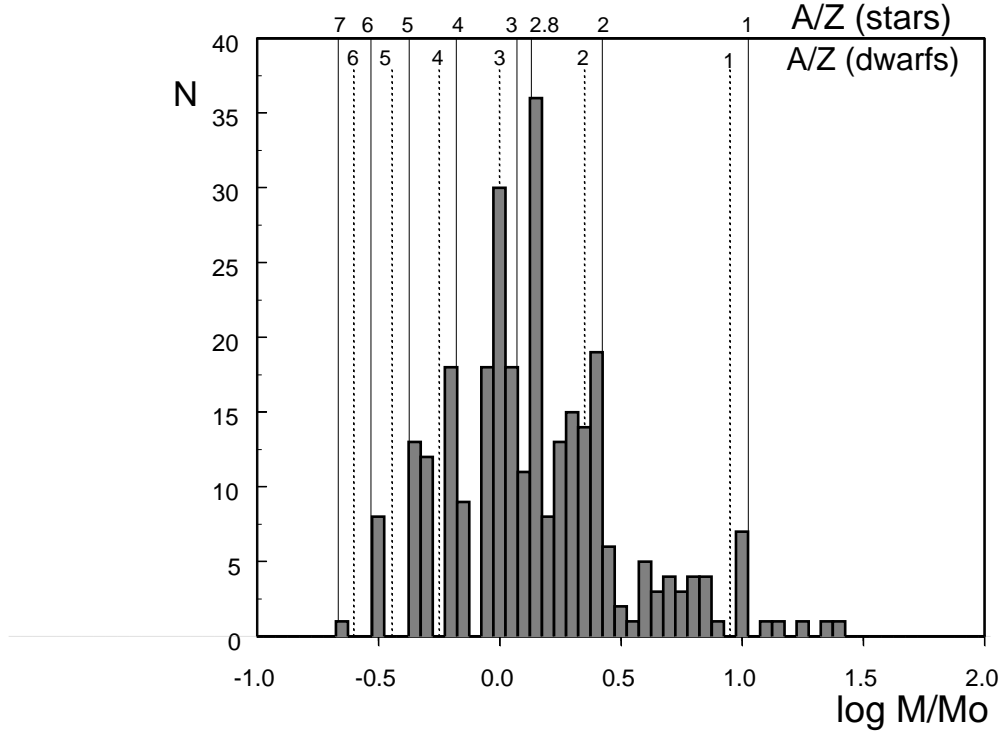}
%\vspace{11cm}
\caption{ The mass distribution of  binary stars \cite{Heintz}. On
abscissa, the logarithm of the star mass over the Sun mass is
shown. Solid lines mark masses  which agree with selected values
of A/Z from Eq.({\ref{4b}}) for stars. The dotted lines mark
masses
 which agree with selected values of A/Z from
Eq.({\ref{5b}}) for dwarfs.} \label{stars}
\end{center}
\end{figure}

The comparison of calculated results with the data of measurements
is shown in Fig.{\ref{stars}}. There is an extensive star mass
evidence but the mass measurement of binary stars has a sufficient
accuracy only. The mass distribution of visual and eclipsing
binary stars \cite{Heintz} is shown in Fig.{\ref{stars}}. On
abscissa, the logarithm of the star mass over the Sun mass is
plotted. Solid lines mark masses which agree with selected values
of A/Z for stars from Eq.({\ref{4b}}). The dotted lines mark A/Z
for dwarfs from Eq.({\ref{5b}}).

\clearpage

According to existing knowledge, the hydrogen inside dwarfs is
fully burnt out. In full agreement with it in Fig.{\ref{stars}}
there are hydrogen stars and there are not dwarfs with $A/Z=1$,
whereas there are both - helium-deuterium stars and dwarfs - with
$A/Z=2$. Attention should be attract to the fact that there is the
peak of stars consisting of heavy nuclei with $A/Z=2.8$ and peak
for dwarfs with $A/Z=3$ where the Sun is placed.

The nuclei with $A/Z > 2.8$ are absent in terrestrial condition.
Evidently a situation for stars composed of nuclei with $A/Z > 3$
is more complicated and it demands a special and more attentive
consideration.

\subsection{The mass of a star consisting of neutron matter}

Dwarfs may be considered as stars where a process of
neutronisation is just beginning. Finally, at nuclear density,
plasma turns into neutron matter.

It is agreed that a pulsar is a star consisting of neutron matter
with some impurity of other particles - electrons and protons.
Evidently, at nuclear density neutrons and protons are
indistinguishable inside pulsars as inside a huge nucleus. The
neutron matter can be considered as some kind of  plasma where
electrons and protons exist in a neutron environment. Under this
condition a very small impurity of electrons and protons (about a
level of $10^{-18}$) is enough to induce a sufficient electric
polarization and to balance the action of gravity.

It should be noted that  neutron matter at nuclear density is a
degenerate substance, if $T\ll 10^{12}K$. It is important that the
neutron gas is not ultra-relativistic at a nuclear density. It is
a relativistic gas only when the Fermi momentum of neutrons is
$p_F\approx m_p c$. For this reason, the equation of pressure
balance  for neutron relativistic gas has a complicated form
\cite{1}:

\begin{equation}
\frac{GM_{pulsar}^2}{6RV}=\frac{m_p^4 c^5}{32 \pi^2
\hbar^3}\biggl(\frac{1}{3} sh\xi
-\frac{8}{3}sh\frac{\xi}{2}+\xi\biggr)
\end{equation}
Where $\xi=4 Arsh\frac{p_F}{mc}$.

As a result, the mass of a pulsar is

\begin{equation}
M_{pulsar}= \biggl(\frac{3}{4}\biggr)^{3}
\biggl(\frac{9\pi}{4}\biggr)^{1/2} M_{Ch} F(\gamma) \label{6a}
\end{equation}
where

\begin{equation}
F(\gamma)=\biggr[\frac{\frac{1}{3}sh\xi-\frac{8}{3}sh\frac{\xi}{2}+\xi}
{\bigl(sh\frac{\xi}{4}\bigr)^4}\biggr]^{3/2} \label{6b}
\end{equation}
and

\begin{equation}
sh\frac{\xi}{4}=\frac{p_F}{m_pc}
=\frac{c~m_p^{4/3}}{\hbar(3\pi^2\gamma)^{1/3}}\label{6c}
\end{equation}
depends on world constants and the density of neutron matter
$\gamma$ only.

\begin{figure}
\begin{center}
\includegraphics[4cm,1cm][10cm,11cm]{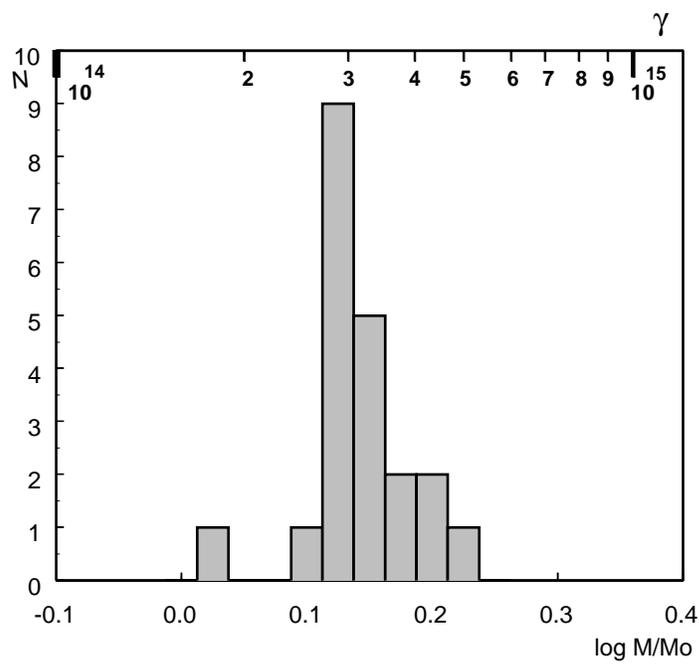}
%\vspace{11cm}
\caption { The mass distribution of pulsars \cite{Thorsett}. In
lower abscissa, the logarithm of the pulsar mass over the Sun mass
is shown. In upper abscissa, the density of the substance in
$g\cdot cm^{-3}$ according to Eqs.({\ref{6a}})-({\ref{6c}}) is
plotted.} \label{pulsar}
\end{center}
\end{figure}

The estimation shows that at $p_F\approx m_p c$ the function
$F(\gamma)\approx 1$. The numerical calculation of $F(\gamma)$
shows that the calculated value of mass of a pulsar
(Eq.({\ref{6a}}))  fully coincides with observed data
(Fig.({\ref{pulsar}})) when the matter density $\gamma$ approaches
 the nuclear density $3\cdot 10^{14}~ g/cm^3$ in full agreement
with the assumption of incompressibility of nuclear matter. (At
this density, $\frac{p_F}{m_p c}\approx 0.4$ and $F(\gamma)\approx
0.72$).

The mass distribution of pulsars is shown in Fig.({\ref{pulsar}}).
In the upper scale the density of neutron matter according to
Eqs.({\ref{6a}})-({\ref{6c}}) is plotted.

\clearpage
\section{Conclusion.}\vspace{12pt}

To conclude, it is important to note that no
disputable speculative assumptions were made above. All above results was
obtained on the standard physical base by standard formal methods.
\vspace{12pt}

The novelty of the developed  approach to plasma equilibrium consists in
the rejection of the usually accepted point of view that an
increase of temperature and density depthward of a celestial
body is a requirement of an equilibrium of a substance in a
gravity field. It is really applicable to  equilibrium of atomic
substances. It is shown above that there is the  minimum for
electron-nuclear plasma energy at  some density and temperature
which brings to the equilibrium in a gravitation field at zero
gradient of the general parameters of plasma  because of its
electrical polarization.\vspace{12pt}

The understanding of this effect discovers the simple mechanism of
generation of the magnetic field by celestial bodies. It can be
noted that all previous models tried to solve the other basic
problem: they tried to calculate the magnetic field of a
celestial body. Now space flights and the development of astronomy
discovered a remarkable and previously unknown fact: the magnetic
moments of all celestial bodies are proportional to their angular
momenta and the proportionality coefficient is determined by the
ratio of world constants only (\cite{Blackett},\cite{Sirag}). The
explanation of this phenomenon is really the basic problem of planetary
and stars magnetism nowadays. The developed theory gives
a simple and standard solution to this problem.\vspace{12pt}

    This approach gives a possibility to predict
important properties of stars in their steady state. Starting from
the equilibrium conditions it allows to calculate masses of
different types of stars. Thus the masses of stars composed by
non-relativistic non-degenerate plasma and dwarfs composed by
relativistic degenerate plasma can be expressed by the ratio of
world constants and one variable parameter $(A/Z)$ only, and this
statement is  in a rather good agreement with the observation data.
Just as the predicted value of mass of pulsars is in full
agreement with observations at the assumption of incompressibility
of nuclear matter. \vspace{12pt}

Some considered questions are described more systematically  in
\cite{BV},\cite{BV-book}.

\end{document}